# Experimental realization of a universal RF photonic integrated circuit

Mehedi Hasan,[1,*] Gazi Mahamud Hasan,[1] and Trevor Hall,[1]

[1]Photonic Technology Laboratory, Centre for Research in Photonics, Advanced Research Complex, University of Ottawa, 25 Templeton Street, Ottawa, K1N 6N5, ON, Canada

*mhasa067@uottawa.ca

**Abstract**

This paper demonstrates the operation of a universal RF photonic circuit that consists of a generalized Mach Zehnder Interferometer architecture with its N × 1 combiner replaced by a N × N optical DFT network. The proposed circuit architecture can be used for SSB modulation, frequency up-conversion/down-conversion, frequency multiplication and phase correlated subcarrier generation based on the choice of output ports and dimension (N) of the circuit. An alternative representation of the circuit is also reported with the aim of a systematic demultiplexing. A transfer matrix approach is used to develop the theoretical predictions which are verified by simulation using industry standard software tool. An integrated circuit based on silicon photonics technology readily available in the laboratory is used for experimental confirmation.

## Introduction

Optical processors containing meshes of Mach Zehnder Interferometers (MZI) when suitably configured can provide arbitrary linear processing functions of utility in a variety of applications from fundamental tests of quantum technologies to signal processing in microwave photonics. To the best of the authors' knowledge, the very first integrated MZI single side band (SSB) modulator was demonstrated in 1981 [1]. The design can be characterized as an instance of a Generalized Mach-Zehnder Interferometer (GMZI) circuit architecture, where an array of N parallel linear electro-optic phase modulators interconnect the egress ports of a 1×N splitter input coupler and the ingress ports of a N×1 combiner output coupler. Since then a plethora of publications [2-11] have applied the same architecture for different RF photonic applications including, but not limited to, SSB modulation, frequency up-conversion/ frequency shifting operation, RF frequency multiplication and phase correlated multiple carrier generation to facilitate high data transmission. The variations in the static optical and electrical phase shifts are the main differences among these circuit architectures. A universal coherent electro-optic circuit architecture is reported in [12], that subsumes all these applications in one circuit. The replacement of the N×1 combiner of a GMZI circuit by an N×N optical Discrete Fourier Transform (DFT) network is the principal innovation. Although the principal application is the generation of N phase-correlated harmonically separated subcarriers for next generation optical transmission systems, the circuit performs other applications reported in the prior art [1-11] given an appropriate choice of output ports and dimension N. The generation of N phase correlated carriers from different output ports eliminates the requirement of an optical de-multiplexing filter needed to separate subcarriers generated from same port by means of optical comb techniques [13-16].

This paper reports the experimental findings obtained from a prototype that validates the operating principles of the universal RF photonic circuit architecture disclosed in [12] for the dimension N = 4. A silicon photonics integrated circuit consisting of a 4 × 4 MMI splitter connected to a 4 × 4 MMI combiner via an array of four parallel dual-drive Mach Zehnder Modulators (MZM) fabricated originally for a frequency up-conversion application [17] is adapted to this experiment. The details of the fabrication technology and electro-optic characterization of the phase shifters can be found in [17]. The eight on-chip electro-optic phase shifters are effectively reduced to the four required for the verification of the proposed RF photonic circuit architecture by the selection of an appropriate drive.

The paper is structured as follows. Firstly, the general form of a universal RF photonic circuit architecture is explained, and an equivalent realization is developed. Secondly, a universal architecture is specialised to dimension $N = 4$. The theoretical analysis of the proposed architecture ($N = 4$) is then developed using a transfer matrix approach. Followed by the theoretical prediction of the adapted photonic integrated circuit under specific operating conditions is developed and verified by computer simulation. Finally, the experimental measurements are reported.

### Principal of operation

Figure 1(a) shows the schematic of the proposed discrete Fourier optics RF photonic circuit. The circuit consists of a $1 \times N$ uniform splitter that excites an array of N electro-optic phase modulators with outputs combined by a $N \times N$ optical DFT network. The optical DFT network is a passive coherent network formed by a combination of 180° hybrids implemented as $2 \times 2$ MMI couplers. The phase modulators are driven by an RF source having angular frequency of $\omega$ with a progressive phase shift of $2\pi/N$. The details of the mathematical derivation are reported in [12]. The analysis reveals that the proposed architecture subsumes all MZI based circuit reported in the prior art given appropriate selection of dimension N and output port(s). Figure 1(b) shows an alternative realization of the circuit architecture drawn in Fig.1(a), where all the parallel phase modulators are replaced by an equivalent delay. The complex optical field over different paths is sampled by a progressive delay of $T/N$, where T is the nominal period of the RF carrier. This is equivalent to sampling the input optical envelope at different times [18]. The merit of this architecture is that it does not require a progressive RF phase shift to drive N phase modulators. However, the fixed sampling interval limits the operating RF bandwidth.

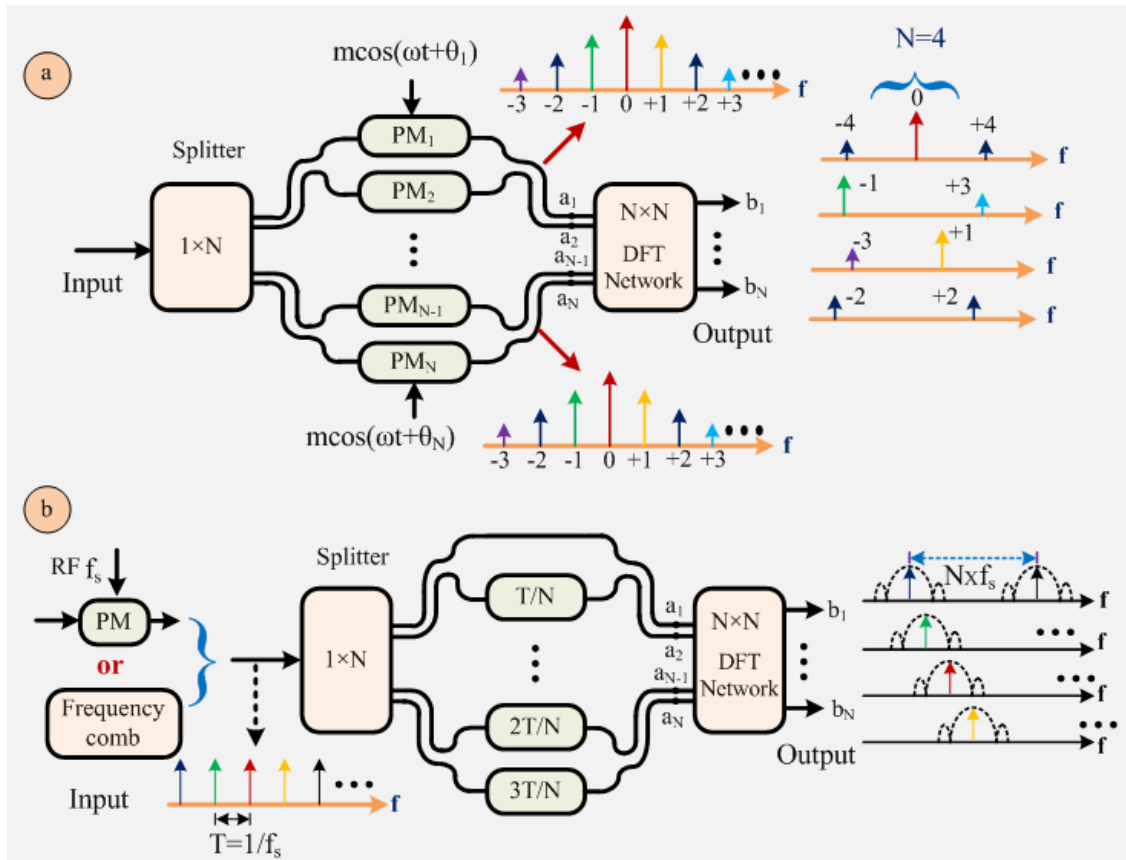

Fig. 1. Fig. 1. (a) Architectural diagram of the proposed universal RF photonic architecture, where an array of N parallel phase modulators are sandwich between $1 \times N$ splitter and $N \times N$ DFT network; (b) An alternative realization of the schematic shown in Fig. (a). PM, phase modulator; DFT, discrete Fourier Transform. The output sketch is drawn for N=4.

Figure 2(a-b) shows the simulated transmission spectra of the circuit for the dimension $N = 2$ and $N = 4$ respectively. A sampling interval of $T = 0.1$ ns is considered in the simulation. Result shows that the adjacent channel cross talk is low only at the channel passband centres. On the one hand, the array of parallel delay line can be mapped into a cascaded delay with a reduced number of arms, thereby reducing circuit complexity [19]. The cascaded architecture can be used as an RF frontend signal processor [20]. On the other hand, the circuit presented in Fig. 1(a) can be operated over a range of frequencies up to the bandwidth of the phase modulator with the aid of an RF drive progressive phase shift circuit that, for example, constructs appropriate linear superpositions of in-phase and quadrature-phase components of the RF source. It is shown in reference [12], that a permutation of the phase modulators can lead to a differentially driven MZM implementation, which renders implicit the relation between the progressive phase shift and the phase modulators. In addition, the complexity of DFT network can be reduced by using higher dimension MMI couplers. For example, a $4 \times 4$ DFT network can be realized by a single $4 \times 4$ MMI coupler with properly chosen phase shifter elements [21] as illustrated in Fig. 3(a). The main difference from the circuit presented in Fig. 1(a) is that the $1 \times N$ uniform splitter is replaced by a $4 \times 4$ MMI splitter, and the ($\times 4$) DFT network is replaced by a $4 \times 4$ MMI coupler augmented by with properly chosen phase shift elements ($\varphi_1$ and $\varphi_2$ in this case). A photonic integrated circuit, readily available in our laboratory, is adapted to verify the operation of RF photonic circuit with a $4 \times 4$ DFT network as shown in Fig. 3(a). Figure 3(b) shows the schematic of the integrated circuit which consists of eight (8) parallel phase shifters with $4 \times 4$ MMI couplers as the input splitter and output combiner. The function of Fig. 3(a) can be emulated by the circuit in Fig. 3(b) using carefully selecting the phase shifter RF drive and bias. The remainder of this section develops an expression for the output of the circuit of Fig, 3(a) using the transfer matrix method. It is also shown theoretically, how the circuit of Fig. 3(b) may emulate the circuit of Fig. 3(a) as verified by simulation using an industry standard software tool.

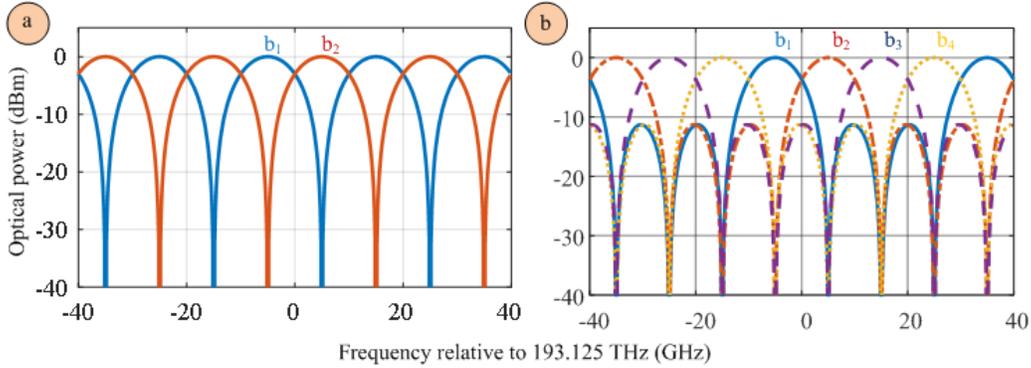

Fig. 2(a). Simulated transmission spectra of the circuit as shown in Fig.1(b) for $N = 2$; (b) transmission spectra of the same circuit when the dimension $N = 4$. The sample interval T is considered as $0.1\ ns$ (10 GHz free-spectral range) in the simulation.

Transmission matrix of the circuit architecture shown in Fig. 3(a) is:

$$\begin{bmatrix} O_1 \\ O_2 \\ O_3 \\ O_4 \end{bmatrix} = T_{4\times 4} \begin{bmatrix} \exp(j\varphi_1) & 0 & 0 & 0 \\ 0 & \exp(j\varphi_2) & 0 & 0 \\ 0 & 0 & -\exp(j\varphi_4) & 0 \\ 0 & 0 & 0 & -\exp(j\varphi_3) \end{bmatrix} T_{4\times 4} \begin{bmatrix} I_1 \\ I_2 \\ I_3 \\ I_4 \end{bmatrix} \qquad (1)$$

where:

$$T_{4\times4} = \frac{1}{\sqrt{4}} \begin{bmatrix} 1 & \zeta & -\zeta & 1 \\ \zeta & 1 & 1 & -\zeta \\ -\zeta & 1 & 1 & \zeta \\ 1 & -\zeta & \zeta & 1 \end{bmatrix} \quad ; \quad \zeta = e^{-i\pi/4} \tag{2}$$

is the transfer matrix of a 4×4 MMI coupler;

$$\varphi_j = \pi v_j / v_\pi \tag{3}$$

is the induced optical phase shift caused by RF signal having amplitude of $v_j$; and $v_\pi$ is the half-wave voltage of the modulator.

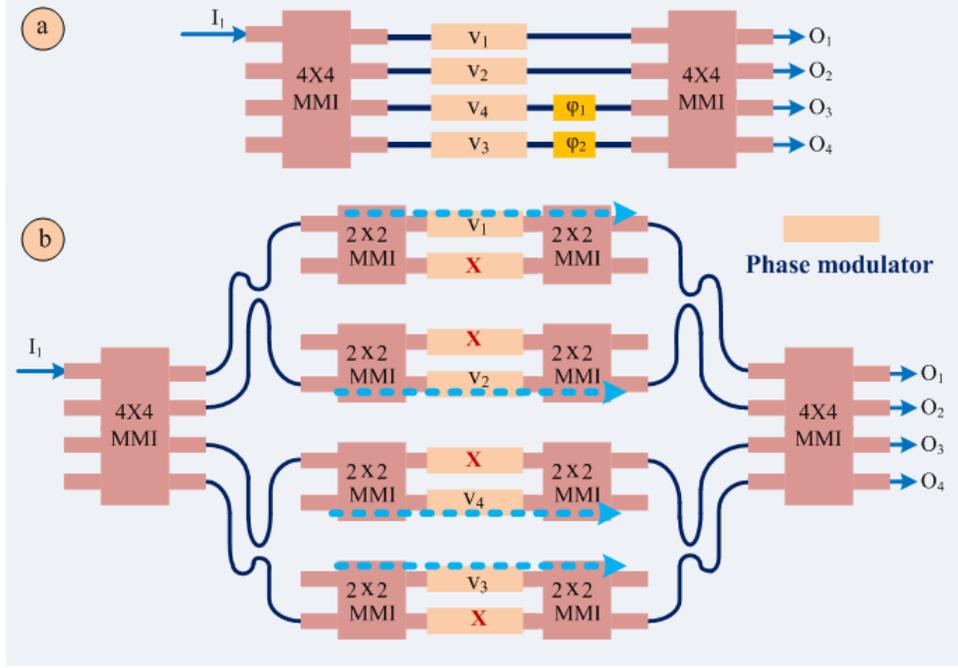

Fig. 3(a) Four parallel phase modulators between $4 \times 4$ optical DFT network and $1 \times 4$ Splitter, thereby create discrete Fourier optics RF photonic circuit [12]; (b) adapted readily available fabricated chip to demonstrate the function of Fig. (a). 'X' represents the unmodulated phase shifter.

If the input $I_1$ is energized only,

$$\begin{bmatrix} O_1 \\ O_2 \\ O_3 \\ O_4 \end{bmatrix} = (1/4)I_1 \begin{bmatrix} \exp(jv_1) - j\exp(jv_2) + j\exp(jv_4) - \exp(jv_3) \\ \zeta\{\exp(jv_1) + \exp(jv_2) + \exp(jv_4) + \exp(jv_3)\} \\ \zeta\{-\exp(jv_1) + \exp(jv_2) + \exp(jv_4) - \exp(jv_3)\} \\ \exp(jv_1) + j\exp(jv_2) - j\exp(jv_4) - \exp(jv_3) \end{bmatrix} \tag{4}$$

In the case of an RF drive of the form:

$$v_1 = m\cos(\omega t); \; v_2 = m\cos(\omega t + \pi/2); \; v_3 = m\cos(\omega t + \pi); \; v_4 = m\cos(\omega t + 3\pi/2) \tag{5}$$

where $m$ is the depth of modulation; defined as;

$$m = \pi v_{RF}/v_\pi \tag{6}$$

After applying the Jacobi Anger expansion and simplifying [22]:

$$\begin{bmatrix} O_1 \\ O_2 \\ O_3 \\ O_4 \end{bmatrix} = I_1 \begin{bmatrix} i\{J_1(m)\exp[j\omega_{RF}t] - J_3(m)\exp[-j3\omega_{RF}t]\ldots\} \\ e^{-i\pi/4}\{J_0(m) + 2J_4(m)\cos(4\omega_{RF}t)\ldots\} \\ e^{-i\pi/4}\{2J_2(m)\cos(2\omega_{RF}t) + 2J_6(m)\cos(6\omega_{RF}t)\ldots\} \\ i\{J_1(m)\exp[-j\omega_{RF}t] - J_3(m)\exp[j3\omega_{RF}t]\ldots\} \end{bmatrix} \qquad (7)$$

It can be observed that the proposed circuit subsumes RF photonic architecture defined in the prior art for single side band suppressed carrier modulation (SSB-SC), I-Q modulation, frequency multiplication and spatially separated carrier generation given an appropriate output port choices and dimension of the phase modulator used. For example, for output port-1 ($O_1$) all the harmonics are suppressed except -$3\omega_{RF}$, +$1\omega_{RF}$; which provides a +$\omega_{RF}$ shift of the optical carrier frequency or equivalently upper SSB modulation. Similarly, output port-4 ($O_4$) provides a $-\omega_{RF}$ shift of the optical carrier frequency or lower SSB modulation. For output port-2 ($O_2$) all the harmonics are suppressed except the carrier and $\pm 4^{th}$ harmonics. For output port-2 ($O_2$) all the harmonics are suppressed except $\pm 2\omega_{RF}$ and $\pm 6\omega_{RF}$. If impinged onto a photodiode, a frequency multiplication of four times ($4\omega_{RF}$) can be obtained. Each of the output of the complete architecture may be used to spatially separate a subcarrier to facilitate high data rate transmission. Table 1 summaries the circuit functions when four phase-modulators are used.

Table 1. Functions provided by circuit architecture presented in Fig. 3(a)

| Function | Port(s) used | Remarks |
| --- | --- | --- |
| IQ modulation, SSB-SC & electro-optic frequency conversion | 1($O_1$) or 4 ($O_4$) | |
| × 4 RF frequency multiplication | 3 ($O_3$) | × 8 RF can be generated from port-2 by suppressing the carrier |
| Spatially separated carrier generation | all | |

For identical optical path lengths, when the upper arm is driven only, the transfer matrix of a $2 \times 2$ MZM can be written as:

$$\begin{bmatrix} O_1 \\ O_2 \end{bmatrix} = \frac{1}{2}\begin{bmatrix} 1 & -i \\ -i & 1 \end{bmatrix}\begin{bmatrix} \exp(j\varphi) & 0 \\ 0 & 1 \end{bmatrix}\begin{bmatrix} 1 & -i \\ -i & 1 \end{bmatrix}\begin{bmatrix} I_1 \\ I_2 \end{bmatrix} \qquad (8)$$

where $\frac{1}{\sqrt{2}}\begin{bmatrix} 1 & -i \\ -i & 1 \end{bmatrix}$ is the transfer matrix of a 2 × 2 MMI. Simplifying Eq. (8)

$$\begin{bmatrix} O_1 \\ O_2 \end{bmatrix} = \frac{1}{2}\begin{bmatrix} \exp(j\varphi) - 1 & -i\{\exp(j\varphi) + 1\} \\ -i\{\exp(j\varphi) + 1\} & 1 - \exp(j\varphi) \end{bmatrix}\begin{bmatrix} I_1 \\ I_2 \end{bmatrix} \qquad (9)$$

Similar way, when the lower arm is driven only. The transfer matrix can be written as,

$$\begin{bmatrix} O_1 \\ O_2 \end{bmatrix} = \frac{1}{2}\begin{bmatrix} 1 - \exp(j\varphi) & -i\{\exp(j\varphi) + 1\} \\ -i\{\exp(j\varphi) + 1\} & \exp(j\varphi) - 1 \end{bmatrix}\begin{bmatrix} I_1 \\ I_2 \end{bmatrix} \qquad (10)$$

In the fabricated chip, only the minimum transmission port of the MZMs are connected to the outer $4 \times 4$ MMI. Hence, one can use either $\exp(j\varphi_1) - 1$ or $1 - \exp(j\varphi_1)$ as the transmission of each MZM based on the input port selection. The transmission matrix of the adapted circuit as shown in Fig. 3(b), when only one arm of each MZI is excited is given by:

$$\begin{bmatrix} O_1 \\ O_2 \\ O_3 \\ O_4 \end{bmatrix} = (1/2) T_{4 \times 4} \begin{bmatrix} \exp(j\varphi_1) - 1 & 0 & 0 & 0 \\ 0 & \exp(j\varphi_2) - 1 & 0 & 0 \\ 0 & 0 & 1 - \exp(j\varphi_4) & 0 \\ 0 & 0 & 0 & 1 - \exp(j\varphi_3) \end{bmatrix} T_{4 \times 4} \begin{bmatrix} I_1 \\ I_2 \\ I_3 \\ I_4 \end{bmatrix} \quad (11)$$

where $\varphi_j$ is the optical phase shift induced by RF drive signal as defined by Eq. (3).

When the light is launched into input $I_1$ only, simplifying Eq. (10) using Eq. (3),

$$\begin{bmatrix} O_1 \\ O_2 \\ O_3 \\ O_4 \end{bmatrix} = (1/8) I_1 \begin{bmatrix} \exp(jv_1) - j\exp(jv_2) + j\exp(jv_4) - \exp(jv_3) \\ \zeta\{-4 + \exp(jv_1) + \exp(jv_2) + \exp(jv_4) + \exp(jv_3)\} \\ \zeta\{-\exp(jv_1) + \exp(jv_2) + \exp(jv_4) - \exp(jv_3)\} \\ \exp(jv_1) + j\exp(jv_2) - j\exp(jv_4) - \exp(jv_3) \end{bmatrix} \quad (12)$$

The applied RF drive is as given by Eq. (5). Applying Jacobi Anger expansion and simplifying,

$$\begin{bmatrix} O_1 \\ O_2 \\ O_3 \\ O_4 \end{bmatrix} = (1/2) I_1 \begin{bmatrix} i\{J_1(m)\exp[j\omega_{RF}t] - J_3(m)\exp[-j3\omega_{RF}t] \dots \} \\ e^{-i\pi/4}\{J_0(m) - 1 + 2J_4(m)\cos(4\omega_{RF}t) \dots \} \\ e^{-i\pi/4}\{2J_2(m)\cos(2\omega_{RF}t) + 2J_6(m)\cos(6\omega_{RF}t) \dots \} \\ i\{J_1(m)\exp[-j\omega_{RF}t] - J_3(m)\exp[j3\omega_{RF}t] \dots \} \end{bmatrix} \quad (13)$$

If $I_2$ is excited,

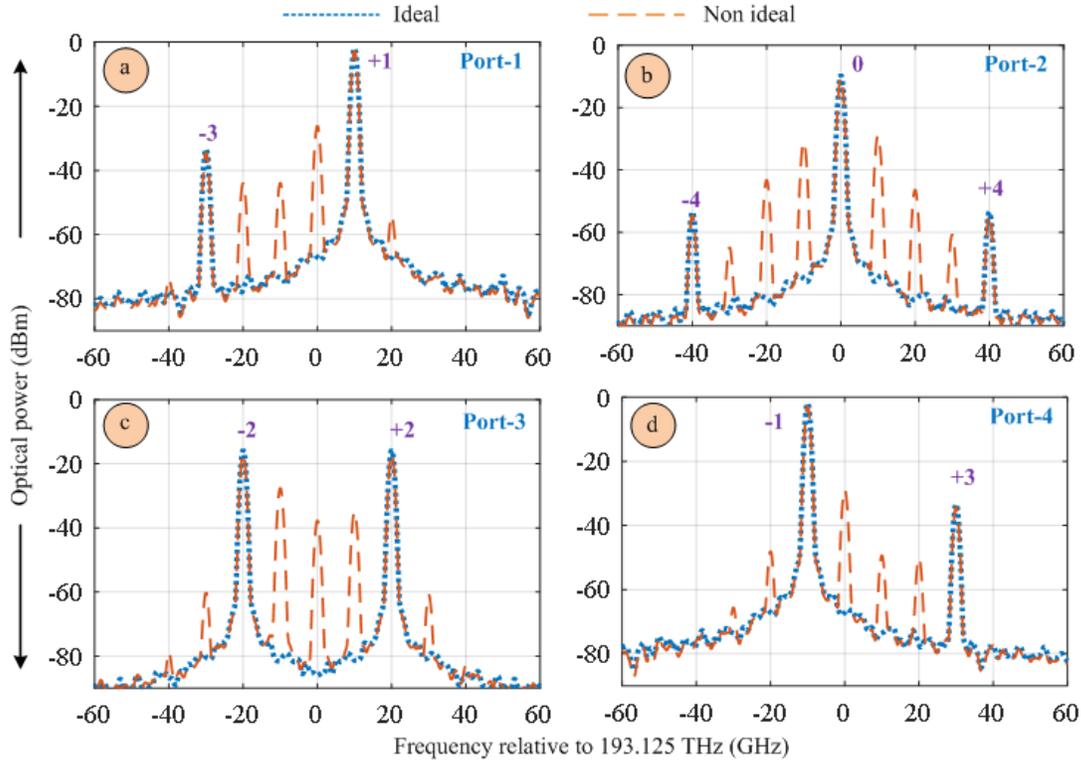

Fig. 4. Simulation result of the schematic presented in Fig. 3(b); (a-d) Optical spectrum at the four output ports (port-1, port-2, port-3 and port-4 respectively) of the $4 \times 4$ MMI when light is launched from the input port-1; The ideal plots are obtained when there no phase error and power imbalances among the ports of MMIs, whereas the non-ideal plots are obtained by adding power imbalances and phase errors in the MMIs according to the Table 2.

$$\begin{bmatrix} O_1 \\ O_2 \\ O_3 \\ O_4 \end{bmatrix} = \frac{1}{2} I_2 \begin{bmatrix} e^{-i\pi/4}\{J_0(m) - 1 + 2J_4(m)\cos(4\omega_{RF}t) \ldots\} \\ i\{J_1(m)\exp[-j\omega_{RF}t] - J_3(m)\exp[j3\omega_{RF}t] \ldots\} \\ i\{J_1(m)\exp[j\omega_{RF}t] - J_3(m)\exp[-j3\omega_{RF}t] \ldots\} \\ e^{-i\pi/4}\{2J_2(m)\cos(2\omega_{RF}t) + 2J_6(m)\cos(6\omega_{RF}t) \ldots\} \end{bmatrix} [I_2] \quad (14)$$

Eq. (13) is identical to Eq. (7) albeit the power at the output of the adopted chip is reduced by -6dB. In addition to, the output at port-2 ($O_2$) contain an extra carrier term contributed by the unmodulated arm of the MZM. Although, the constant carrier term is a part of the output of all the MZMs, it cancels out automatically from all the output ports due to inherent input/output phase relationship of the adapted circuit except from port $O_2$.

To verify this theoretical prediction, a simulation is performed using the industry standard software tool, VPIphotonics. A CW-DFB laser having output power of 20 mW with an operating vacuum wavelength of 1550 nm is used as an optical source. A 10 GHz RF signal having peak amplitude of $0.25v_\pi$ and appropriate phase shift as per Eq. (4) is applied to the phase modulators. Figure 4. (i-iv) shows the optical spectrum at the output ports of the $4 \times 4$ MMI. The output spectrum justifies the theoretical prediction as derived in Eq. (13).

**Table 2. Summary of the power imbalances and phase error of the MMIs [23]**

| Type of MMI | Output ports | Power splitting ratio | Phase errors |
|---|---|---|---|
| 4×4 | 1 | 25% | 2° phase error is considered between output ports 1 & 4 |
|  | 4 | 23% |  |
|  | 2 | 24% | 2° phase error is considered between output ports 2 & 3 |
|  | 3 | 22% |  |
| 2×2 | 1 | 50% | 1° phase error |
|  | 2 | 49% |  |

In practice, the power imbalances and phase errors between the ports of an MMI caused by variation in the fabrication processes deviate the correct operation of the circuit from ideal behaviour. The S-matrix of the MMI is perturbed from ideal to account the fabrication tolerances. In this simulation, the S-matrix of all the $2 \times 2$ MMI along with the two $4 \times 4$ MMI are configured according to the parameters listed in the Table 2 and loaded in VPIphotonics. The details about MMI tolerances can be found in [31]. The MMI design can be improved further using sub-wavelength engineering and broadband operation can be assured [24]. Simulated optical spectra shows that strong carrier breakthrough occurs at the output of the circuit. To improve the performance an integrated variable optical attenuator (VOA) [25] can be used to improve extinction or, to be more systematic, a trimming algorithm applied [26].

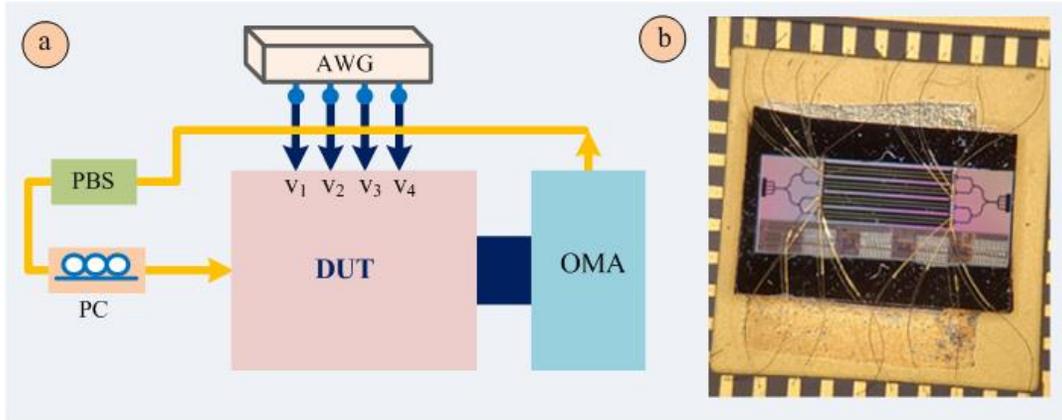

Fig. 5. (a) Schematic of the measurement setup; (b) microscopic image of the wire bonded chip; AWG, arbitrary waveform generator; PBS, polarization beam splitter; PC, polarization controller; OMA, optical modulation analyzer; DUT, device under test.

**Results and discussion**

A photonic integrated circuit readily available in the laboratory, fabricated for another application is adapted for the experimental demonstration of a discrete Fourier transform based RF photonic integrated circuit. Figure 5 shows the photograph of the fabricated chip with wire bonding. The device was fabricated using CMOS compatible SOI technology. The thickness of the buried oxide layer was 2 $\mu m$, whereas the thickness of the top Si layer was 220 $nm$. The wide width of the p-n junction-based phase shifter in comparison to the width of the input/output ports of the MMI introduces waveguide bends into the circuit layout. As such, path length matching becomes a most important and critical aspect of the design. More detail about the fabricated chip can be found in [17]. The detail design of the MMI presented in the fabricated chip can be found in [27].

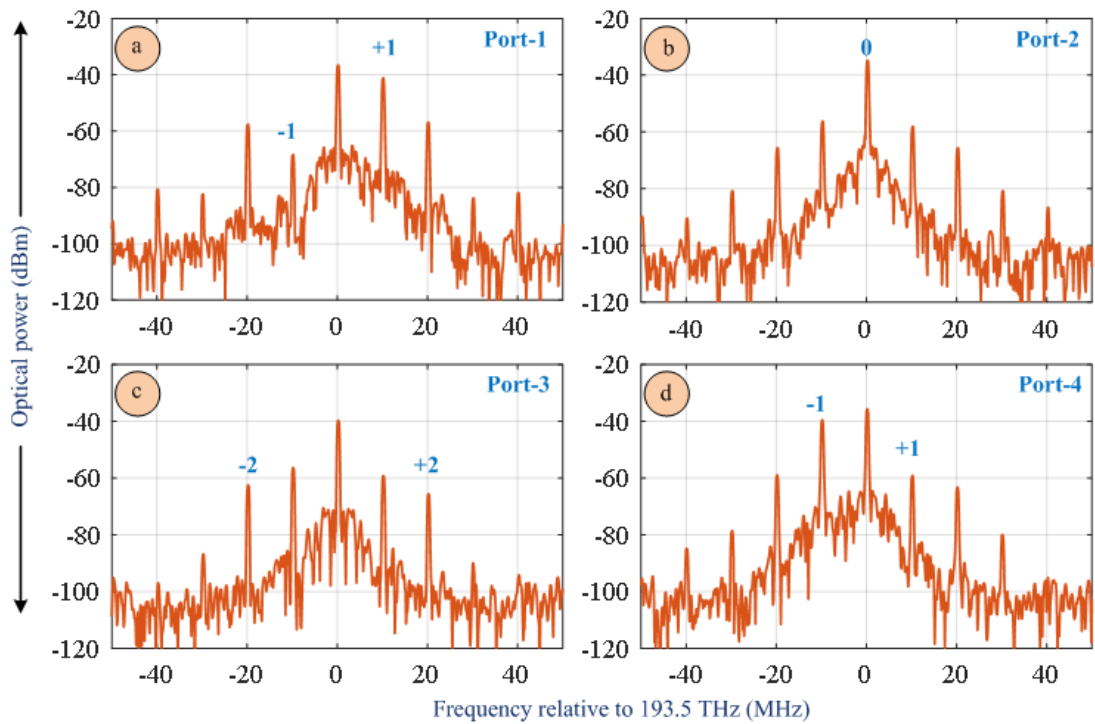

Fig. 6. Measured optical spectra at different output ports of the 4 × 4 MMI when the light is launched from input port 1 ($I_1$); (a) port-1 ($O_1$); (b) port-2 ($O_2$); (c) port-3 ($O_3$); (d) port-4 ($O_4$).

A continuous wave laser at a wavelength of 1550.38 nm is used as the optical input. A polarization beam-splitter (PBS) followed by a polarization controller (PC) is used to constrain the polarization of the light input to the silicon chip to be TE-like. The output of the circuit is connected to an optical modulation analyzer (Agilent N4391A) which is connected to a digital storage oscilloscope (Infiniium DSO-20-91604A). The N4391A is an Optical Modulation Analyzer (OMA). An OMA is a coherent receiver that converts a signal to baseband I & Q components using a local oscillator – i.e., the I & Q components are the real and imaginary parts of the complex envelope of a modulated carrier (at the LO frequency) that represents the signal. The Fourier transform of the baseband envelope consequently provides the spectrum of the signal centered at the LO frequency. To maintain the same polarization state, polarization maintaining fiber is used for all connections. Lensed fibers are used for fiber-to-chip coupling at the input-output to reduce coupling loss. The four phase modulators are driven by four RF signals with specific phase relationship generated by a four-channel arbitrary waveform generator (AWG) (Fluke 294). The four RF signals with progressive phase shifts are denoted as $V_1$, $V_2$, $V_3$ and $V_4$ in Fig 5. A 10 MHz sinusoid signal with amplitude 2.5V is applied to each phase shifter. The RF phase relationship among the channels are consistent as per the settings around 10 MHz frequency even though the maximum frequency limit was 40 MHz. For this circuit to be operated as a DFT network, careful positioning of the RF drives from the AWG is necessary. For example, to maintain symmetry with the ideal DFT circuit shown in Fig. 3(a), RF drives $V_4$ and $V_3$ are applied to the phase modulators at the lower and upper arm of the $MZM_3$ and $MZM_4$ respectively so that the π optical phase shifts can be naturally derived from the phase relationship of the input-output ports of the corresponding 2×2 MMIs. A DC voltage of -3 V is applied via bias tee to make sure the p-n junction always operates in reverse bias. The DC biases are tuned slightly to obtain the correct operation. Light from the optical source is applied to individual input of the chip to investigate the response separately. The measured optical spectra of the four output ports of the circuit when the optical source is connected to input 1 of the circuit is shown in Fig. 6. Output port-1 and port-4 generates upper sideband and lower sideband respectively aligned with the theoretical prediction. The theoretically pristine performance of the circuit is limited by the presence of carrier and undesired spurious sidebands.

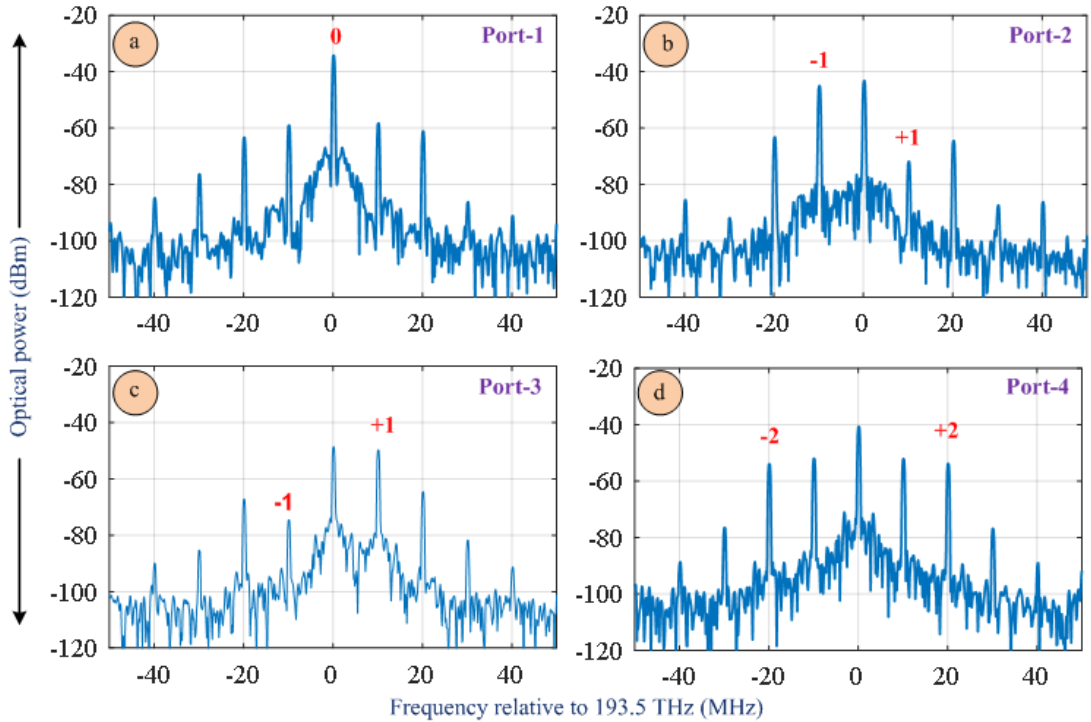

Fig. 7. Measured optical spectra at different output ports of the 4 × 4 MMI when the light is launched from input port 2 ($I_2$); (a) port-1 ($O_1$); (b) port-2 ($O_2$); (c) port-3 ($O_3$); (d) port-4 ($O_4$).

The carrier breakthrough is too high due to power and phase imbalances between the ports of the MZM and the inactive phase shifter arm in each of the MZM working as bypass light path. Furthermore, the modulators are driven at a peak voltage of 2.5V due to limitation of the source, resulting in very small modulation index so that a suppressed carrier amplitude remains strong in comparison to the weak sidebands. The output spectrum at port-2 well justifies the theoretical prediction. Output port-3 should result in the dominance of the +2 and -2 harmonics. The +1 and -1 harmonics are suppressed to -60 dBm from circa -40 dBm. Due to the small modulation index, the amplitude of the $2^{nd}$ harmonic is too low. High insertion loss is also observed. The principal reason is that the chip lacks any mechanism for matching between the fiber mode and high-contrast access waveguide mode and some defects have also been observed on the chip [17].

To verify the correct operation of the circuit, further investigation is performed by changing the input light injection from input port-1 to port-2. The output spectrum is shown in Fig. 7. Now the optical carrier and $\pm 2^{nd}$ harmonic excite from the port-1($O_1$) and port-4 ($O_4$) respectively. Similarly, output port-2 and port-3 generates LSB and USB respectively. The same kind of deviations from the theoretical prediction are observed as shown in Fig. 7. Nevertheless, it can be concluded that the circuit functions as universal RF photonic circuit described in [12] which subsumes the majority of the prior art in RF photonics. The main aim of this report is to prove that the fabrication of $N = 4$ parallel phase modulators combined by an $(4 \times 4)$ DFT circuit can be made practical. Careful design with incorporating tuning mechanism via means of a phase shifter, VOA [25] or more in a systematic way, i.e. by using algorithm [26], can be applied to improve further the performance of the circuit.

**Conclusion**

In summary, a previously proposed RF photonic circuit architecture consisting of a $1 \times N$ splitter connected with $N \times N$ DFT network via N parallel electro-optic phase modulator has been implemented. A photonic integrated circuit based on silicon photonics technology having eight parallel phase shifters connected between $4 \times 4$ MMI at the input and output is configured for the target application. All though phase-correlated but spatially separated subcarrier generation is the principal aim, the circuit can subsume majority of the RF photonic applications in the prior art. Experimental results verify the theoretical prediction but the unwanted harmonics suppression ratio and specifically the carrier suppression is not sufficient for two reasons. Firstly, due to the unavailability of high-power RF signal source having four output channels, the phase shifters are driven at low modulation index. Consequently, the $\pm 1^{st}$ harmonics are weak compared to the carrier so that magnitude of the carrier breakthrough due to fabrication tolerances and bias errors remains significant relative to the magnitude of the sidebands. There is no tuning element in the circuit to compensate the MMI output power imbalances and phase errors due to the fabrication and bias errors. Nevertheless, it has been demonstrated that the universal RF photonic architecture described in [12] can be materialized using current state of the art technology, thereby, establishing a systematic design approach for future RF photonic applications.